\documentclass[showpacs,amsmath,12pt,prd,onecolumn,]{revtex4}
\usepackage{graphicx}
\usepackage{latexsym}
\usepackage{epsfig}
\usepackage{amssymb}
\usepackage{color}
\usepackage{hyperref}
\hypersetup{colorlinks,citecolor=blue}
\usepackage{multirow}

\begin{document}
\preprint{}

\title{Possible existence of stable compact stars  in $\kappa(\mathcal{R},\mathcal{T})-$gravity}

\author{G R  P  Teruel}
\email{gines.landau@gmail.com}
\affiliation{Departamento de Matemáticas, IES Carrús, Elche 03205, Alicante, Spain}

\author{Ksh. Newton Singh}
\email{ntnphy@gmail.com}
\affiliation{Department of Physics, National Defence Academy, Khadakwasla, Pune-411023, India}

\author{Farook Rahaman}
\email{rahaman@associates.iucaa.in}
\affiliation{Department of Mathematics, Jadavpur University, Kolkata 700032, West Bengal,
India}

\author{Tanmoy chowdhury}
\email{tanmoych.ju@gmail.com}
\affiliation{Department of Mathematics, Jadavpur University, Kolkata 700032, West Bengal, India}

\date{\today}
\begin{abstract}
We present the first interior solutions representing compact stars in $\kappa(\mathcal{R},\mathcal{T})$ gravity, by solving the modified field equations in isotropic coordinates. Further, we have assumed the metric potentials in Schwarzschild's form and a few parameters along with the isotropic condition of pressure. For solving, we use specific choice of the running gravitational constant as $\kappa(\mathcal{R},\mathcal{T})=8\pi-\lambda \mathcal{T} ~~(G=\tilde{c}=1)$. Once arrived at the reduced field equations, we investigate two solutions with $c=1$ and $c \neq 1$, where $c$ denotes here another constant that should not be confused with the speed of light. Then, we investigate each solution by determining the thermodynamics variable {\it viz} pressure, density, speed of sound, and adiabatic index. We found that these solutions satisfy the Bondi criterion, causality condition, and energy conditions. We also found that the $M-R$ curves generated from these solutions satisfy the stringent constraints provided by the gravitational wave observations due to the neutron star merger GW 170817.
\end{abstract}

\maketitle

%\tableofcontents

\section{\label{sec1} Introduction}

The last advances in astrophysical observations have led to a wide interest in the study of the composition of astrophysical compact objects. Moreover, these objects contain ultra-dense nuclear matter in their interiors, which makes them excellent physical laboratories to test possible departures from general relativity (GR). Traditionally the term compact objects or compact stars refers collectively to white dwarfs, neutron stars, and black holes. Compact stars are also called the stellar remnants, as they are often the endpoints of catastrophic events such as supernova explosions and binary stellar collisions. The state and type of a stellar remnant depend mainly on the properties and composition of the dense matter of the star. However, due to the lack of knowledge of the extreme condition and its complex composition it is a formidable task to determine the exact equation of state (EoS) to represent compact stars. Several astrophysical observations measure masses and radii of compact stars \cite{po02,dr02,wa02,co02}, which in turn, constrain the EoS of dense matter in compact stars. The observation of 2 solar-mass neutron stars \cite{de10,an13} indicates that the EoS for such objects should be sufficiently stiffer than the ordinary nuclear matter to accommodate the large mass. This fact enables us to think about the possibility of stable mass configurations with an interior composed of exotic matter to some extent. Even in the case of low mass compact stars, the density of the core matter is much larger than the normal matter. Due to the extreme density, the nuclear matter may consist, apart from ordinary nucleons and leptons, of exotic components in their different forms and phases, such as Bose-Einstein condensates of strange mesons \cite{ka86,ne87,gl98,ba03}, baryon resonances and hyperons \cite{gl97}, as well as strange quark matter \cite{pr97,fa84,sc00}. Applying the embedding class one technique, many authors have explored well-behaved solutions \cite{ntn1,ntn2,mau1,mau2,gad1,gad2}.

The task of building the EoS of matter beyond the nuclear saturation density, important for the description of compact stars, is an active and vast field of research. For a given EoS, the study of physical features of relativistic compact objects in GR is done by obtaining analytic solutions for static Einstein's field equations and then imposing conditions for physical viability. Due to the non-linear character of Einstein's field equations, this usually represents a challenging and complicated task. Since the first exact solution of GR field equations was found by Schwarzschild \cite{sch16}, the amount of exact analytic solutions has been increasing and are extensively used in the studies of neutron stars and black hole formations, both in GR and also in the modified gravity scheme, which includes GR as a particular case in the appropriate limit \cite{ra03,fe99}.
To model relativistic fluids from a more realistic point of view, then one should include Buchdahl \cite{buc67} and Tolman VII \cite{to39, rag15,new20} solutions.

It is well known that the Tolman–Oppenheimer–Volkoff (TOV) equation \cite{tol34,tol39,Oppen39} constrains the structure of a spherically symmetric body of isotropic material which is in static gravitational equilibrium, as modelled by GR. Isotropic and anisotropic compact stars models have also been explored in the framework of modified gravity. In particular, in the context of the Lagrangian $f(\mathcal{R},\mathcal{T})$ theory, Sharif et al. \cite {sha14} have discussed the stability of isotropic self-gravitating stars, while compact solutions with conformal Killing vectors have been studied by Rahaman et al. \cite{co16}. Regarding the anisotropic case, also in the scheme of $f(\mathcal{R},\mathcal T)$ gravity theory, Biswas and his collaborators \cite{bish19} established a new model for a highly anisotropic star system based on a previous model of metric potentials due to Krori-Barua \cite{kro75}. A charged star system supported by Chaplygin EoS was explored by Bhar \cite{bhar20}.  

Recently, one of us \cite{gr18} has proposed  a new modified theory named as $\kappa(\mathcal{R},\mathcal{T})$ gravity. This modified theory,  instead of using a standard modified theory of gravity approach,  is inspired by  Maxwell's and Einstein's original approaches of adding new possible source terms directly to the field equations. We explicitly refer here to the addition of the displacement current term by Maxwell to complete the Electromagnetic field equations, and to the incorporation of the key trace term, $-\frac{1}{2} \mathcal{R}g^{\mu\nu}$ by Einstein to complete the GR field equations. Indeed, it is interesting to note that, though the variational method is a major tool to build a physical theory and its possible generalizations, it should not be placed on an equal footing with other truly foundational principles such as the equivalence principle and the principle of general covariance, which were the two foundational principles of GR, while the variational principle (the Einstein-Hilbert action) was derived and incorporated to the theory later on after the first correct derivation of GR \cite {hil15,ren07}.

In addition, the huge amount of possible modified gravity Lagrangian theories available in the literature seems to suggest that some other foundational principle is lacking beyond the equivalence principle and the principle of general covariance, and the fact that there is no reason to believe that ordinary symmetries and standard conservation laws will always be present in a final theory of Nature, lead us to think that the Non-Lagrangian approach also deserves to be investigated. In this sense, examples of Non-Lagrangian theories can be found in other branches of theoretical physics, such as quantum field theory (QFT), where it is being increasingly clear that these theories populate much of the QFT landscape and also offer new opportunities in the search of new types of 4-manifold invariants \cite{gu17,gadd15}. In light of the above, the study of viable Non-Lagrangian gravity theories deserves consideration as a legitimate alternative to the standard variational approach.\\
Non-Lagrangian $\kappa(\mathcal{R},\mathcal {T})$ gravity theory is a relatively unknown and still not well-explored proposal. However, in the last few years some works have been devoted to studying its cosmological implications \cite{ahm22,arc22}, collecting results that seem compatible with observational data. A model of wormhole in the context of $\kappa(\mathcal{R},\mathcal{T})$ theory was discussed recently by S. Sarkar et al. \cite{sark19}.\\
 So far, no one has investigated compact stars solutions in the framework of $\kappa(\mathcal{R},\mathcal{T})$ theory of gravity, hence it would be interesting to study whether this Non-Lagrangian theory can support acceptable compact star solutions from a physical point of view. This is the purpose of this work. The paper is organized as follows: In Sec.\ref{sec.II}, we have described Einstein's field equations in the scheme of this modified gravity theory. Isotropic coordinates and the line element chosen to solve the equations are presented in Sec.\ref{sec.III}, underlining the differences between the $c=1$ and the $c\neq 1$ cases. The required boundary conditions to match the internal solution with the external vacuum solution are discussed in Sec.\ref{sec.IV}. On the other hand, Sec.\ref{sec.V} deals with the physical acceptability of the obtained solutions, discussing in detail the role of the energy conditions as well as other constraints. The mass-radius relationship is analyzed in Sec.\ref{sec.VI}. Finally, the discussion has been made in Sec.\ref{sec.VII}.
 
 \section{Einstein's field equations in $\kappa(\mathcal{R},\mathcal{T})$ gravity }\label{sec.II}
The field equations in $\kappa(\mathcal{R},\mathcal{T})$ modified gravity are obtained by adding new possible source terms directly to GR field equations as \cite{gr18}
\begin{equation}
R_{ij}-\frac{1}{2} \mathcal{R}\, g_{ij}-\Lambda g_{ij}=\kappa(\mathcal{R},\mathcal{T})\,T_{ij}\label{eq}
\end{equation}
where $g_{ij}$ is the metric potential, $R_{ij}$ is the Ricci tensor, $\Lambda$ is a cosmological constant, $T_{ij}$ is the energy-momentum tensor of the matter source, and $\kappa(\mathcal{R},\mathcal{T})$ corresponds to the Einstein gravitational constant and it is proposed as a function of the traces $\mathcal{T} = g_{ij}T^{ij}$, and Ricci scalar $\mathcal{R} = g_{ij}R^{ij}$. Clearly, the gravitational constant $\kappa$ depends on the scalars, so we can explore the possibility of a varying gravitational constant, i.e. generalization of the original Einstein's gravitational constant (not at the level of an action functional). A varying gravitational constant in the action leads to a Brans-Dicke scalar-tensor theory type \cite{cb61, cb05} with entirely different field equations from Eq. (\ref{eq}). Since the left hand side of the field Eq. (\ref{eq}) is divergence free, we have
\begin{eqnarray}
\label{m1e}\nabla^j \Big[\kappa(\mathcal{R},\mathcal{T}) T_{ij}\Big]=0
\end{eqnarray}
Then, these field equations imply the non-covariant conservation of $T_{ij}$ that can be expressed as
\begin{eqnarray}
\label{m1f}\nabla^j T_{ij}=-\frac{\nabla^j\kappa(\mathcal{R},\mathcal{T})}{\kappa(\mathcal{R},\mathcal{T})}T_{ij}.
\end{eqnarray}
For $\kappa(\mathcal{R},\mathcal{T})\neq 0$.  At very high energies, for some specific choices of the running gravitational constant such as, for example,  $\kappa(\mathcal{R},\mathcal{T})=\kappa(\mathcal{T})=8\pi G -\lambda \mathcal{T}$, we could have that $\kappa(\mathcal{R},\mathcal{T})= 0$; this may imply an exponential expansion in the early universe ruled by a cosmological constant. Examples of other non-conservative theories include Rastall proposal \cite{ras72}, or the more recent Lagrangian $f(\mathcal{R},\mathcal{T})$ theory of gravity by Harko and his collaborators \cite{har11}.\\
Teruel \cite{gr18} has proposed and analyzed some cosmological implications of two particular models. The first one is proposed by setting, $\kappa(\mathcal{T}) = 8\pi G -\lambda \mathcal{T}$, and corresponds to a matter-matter coupling. The second model is characterized by a gravitational constant that varies as $\kappa'(\mathcal{R}) = 8\pi G +\xi \mathcal{R}$, which will provide a coupling between matter and curvature terms.
The choice of the minus sign in the expression for the running gravitational constant $\kappa(\mathcal{R},\mathcal{T})=8\pi G -\lambda \mathcal{T}$, ($\lambda>0$), is motivated for cosmological reasons. Indeed, one can show that the modified Friedmann equations for the opposite choice, i.e, $\kappa(\mathcal{R},\mathcal{T})=8\pi G+\lambda \mathcal{T}$ implies that at sufficiently high densities $H^{2}\sim \rho^{2}$, where $H$ is the Hubble constant. That behavior is worst in terms of divergences than the GR case. \cite{gr18,olm18}.
In the next section, we will proceed to solve the field equations of $\kappa(\mathcal{R},\mathcal{T})$ gravity theory for a specific (isotropic) line element that represents the interior of a compact object assuming the stress-energy tensor of a perfect fluid as the matter sources.

\section{Isotropic Star  Solution}\label{sec.III}
To obtain the exact solutions of Einstein's field equation in the framework of $\kappa(\mathcal{R},\mathcal{T})$ gravity we can proceed as in ref \cite{fr20}. First, we assume that the static spherically symmetric uncharged matter distribution corresponds to the isotropic line element given by
\begin{equation}\label{isotropic}
ds^{2}=e^{\nu}dt^{2}-e^{\mu}(dr^{2}+r^{2}d\theta^{2}+r^{2}\sin^{2}\theta \,d\phi^{2}),
\end{equation}
where $\nu$ and $\mu$ are functions of the radial coordinate $r$ only. For the specific choice of the running gravitational constant given by $\kappa(\mathcal{R},\mathcal{T})=8\pi-\lambda \mathcal{T} ~~(G=\tilde{c}=1)$, we find that the field equations can be written as
\begin{eqnarray}
e^{-\mu}\Big(\frac{{\mu^{\prime}}^{2}}{4}+\frac{\mu^{\prime}\nu^{\prime}}{2}+\frac{\mu^{\prime}+\nu^{\prime}}{r}\Big)=8\pi p(r)\big[1-\alpha\{\rho(r)-3p(r)\}\big], \label{doce}\\
e^{-\mu}\Big(\frac{\mu^{\prime \prime}}{2}+\frac{\nu^{\prime\prime}}{2}+\frac{{\nu^{\prime}}^{2}}{4}+\frac{\mu^{\prime}+\nu^{\prime}}{2r}\Big)=8\pi p(r) \big[1-\alpha\{\rho(r)-3p(r)\}\big], \label{trece}\\
-e^{-\mu}\Big(\mu^{\prime\prime}+\frac{{\mu^{\prime}}^{2}}{4}+\frac{2\mu^{\prime}}{r}\Big)=8\pi\rho(r)\big[1-\alpha\{\rho(r)-3p(r)\}\big].
\end{eqnarray}
In the above equations, $\alpha=\lambda/8\pi$ is a real constant different from zero, $\rho$ and $p$ are the matter-energy density and pressure, respectively, and the prime denotes derivative with respect to radial coordinate only. Under the isotropic condition Eq. (\ref{doce}) must be equal to Eq. (\ref{trece}), we obtain a differential equation by assuming $e^{\nu}=A\psi(r)^{-a}$ and $e^{\mu}=B\psi(r)^{b}$ in the form
\begin{equation}
\psi^{\prime\prime}-c~\frac{{\psi^{\prime}}^{2}}{\psi}-\frac{\psi^{\prime}}{r}=0,
\end{equation}
where we have defined certain constant $c$ as 
\begin{equation}
c=\frac{\frac{1}{2}b^{2}-\frac{1}{2}a^{2}-ab+b-a}{b-a}
\end{equation}
With $a\neq b$. All the parameters $A,B,a,b$ above are real constants. Analytical solutions can be found by setting different conditions for the parameters.

\subsection{The case for $c=1$}
\begin{figure}
\includegraphics[width=0.45\textwidth]{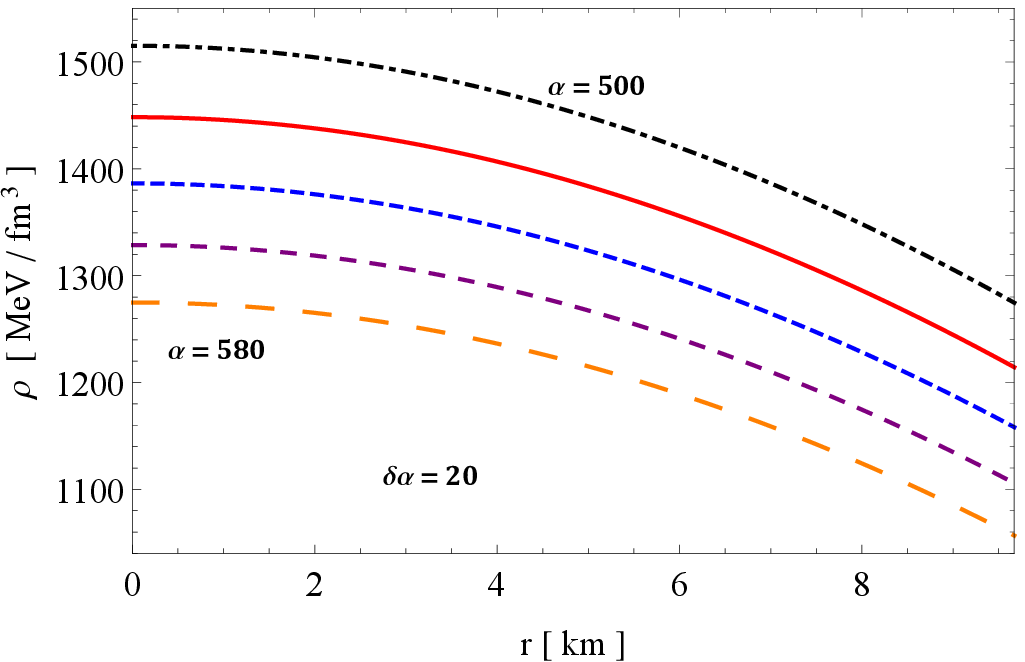}
\hspace{\fill}
\includegraphics[width=0.45\textwidth]{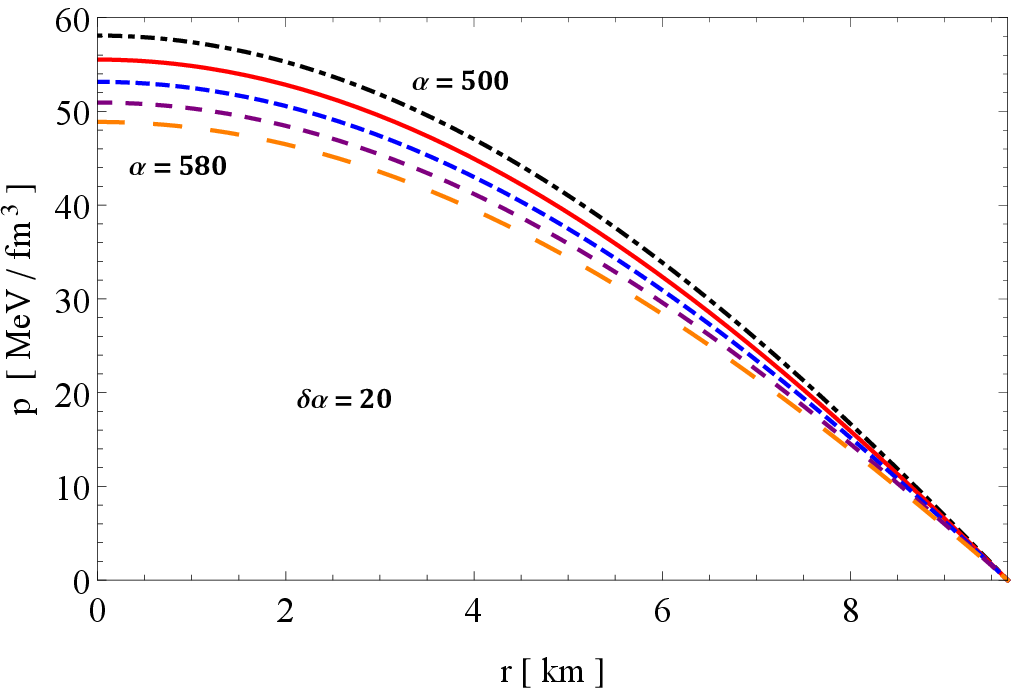}
\caption{Density and Pressure for case A.}\label{fig1}
\end{figure}

\begin{figure}
\includegraphics[width=0.45\textwidth]{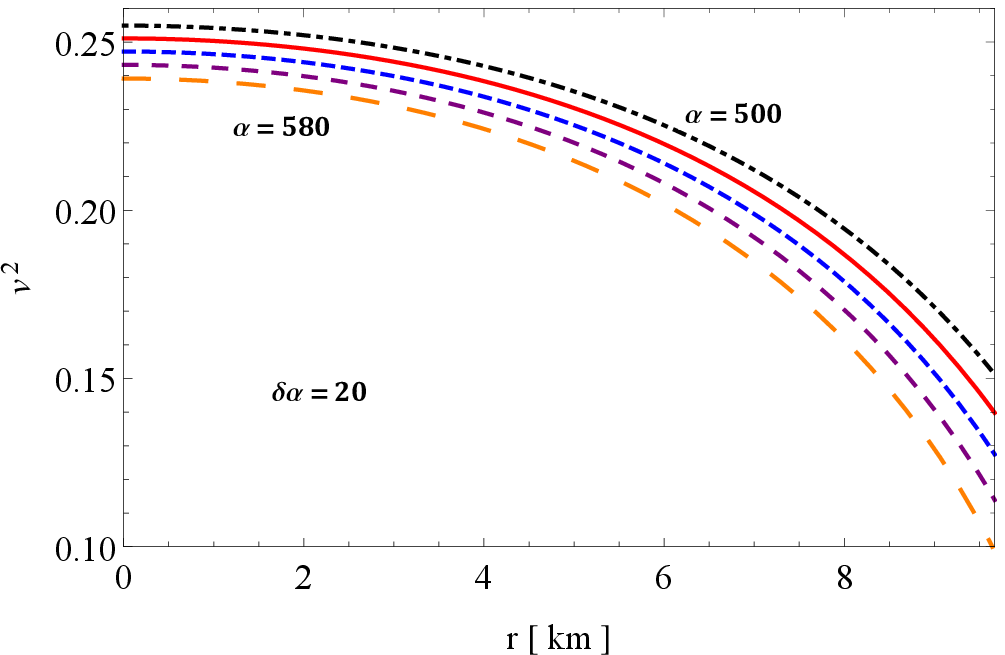}
\hspace{\fill}
\includegraphics[width=0.45\textwidth]{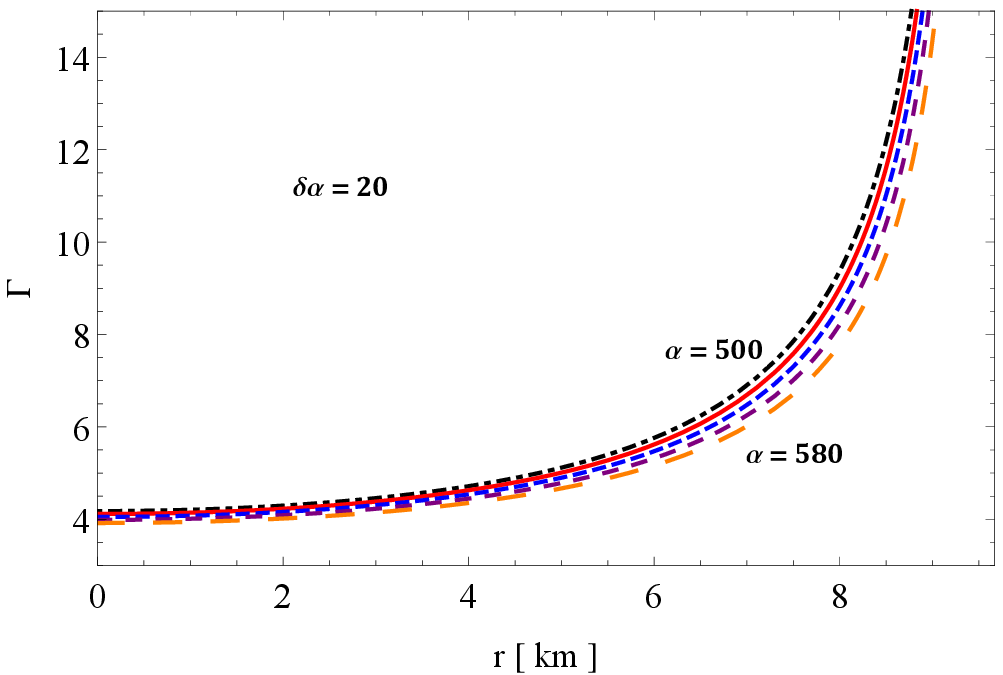}
\caption{Speed of sound and adiabatic index for case A.}\label{fig2}
\end{figure}

For such specific value we obtain the following solutions
\begin{eqnarray}
\psi(r) &=& C_{1}e^{C_{0}r^{2}}~~,~~
e^{\nu} = A_1 e^{-a_1r^{2}}~~,~~
e^{\mu}= B_1 e^{b_1 r^{2}}.
\end{eqnarray}
Here $A_1=AC_{1}^{-a}$, $B_1=BC_{1}^{b}$, $a_1=aC_{0}$ and $b_1=bC_{0}$. The corresponding physical quantities are given below:
\begin{eqnarray}
\frac{C_{1}^{-b}}{B}\Big[b(b-2a)C_{0}^{2}r^{2}+2(b-a)C_{0}\Big]e^{-bC_{0}r^{2}}=8\pi p(r) \big[1-\alpha\{\rho(r)-3p(r)\}\big], \\
-\frac{C_{1}^{-b}}{B}\Big[6bC_{0}+b^{2}C_{0}^{2}r^{2}\Big]e^{-bC_{0}r^{2}}=8\pi\rho(r)\big[1-\alpha\{\rho(r)-3p(r)\}\big].
\end{eqnarray}
The variations of the energy density and the pressure are shown in Fig. \ref{fig1}. From the above equations we get
\begin{equation}
\frac{\rho(r)}{p(r)}=\frac{b(6+bC_{0}r^{2})}{b(2a-b)C_{0}r^{2}+2(a-b)}.
\end{equation}
The speed of sound and the adiabatic index can be determine as
\begin{eqnarray}
v^2 = {dp \over d\rho}~~;~~\Gamma={\rho+p \over p}\, {dp \over d\rho}.
\end{eqnarray}
These variations with respect to radial coordinates are shown in Fig. \ref{fig2}.

\subsection{The case for $c \neq 1$}

For this particular choice we have
\begin{equation}
\psi(r)^{1-c}=C_{0}r^{2}+C_{1}~~,~~
e^{\nu}=A\Big(C_{0}r^{2}+C_{1}\Big)^{-a_1}~~,~~
e^{\mu}=B\Big(C_{0}r^{2}+C_{1}\Big)^{b_1}
\end{equation}
with $a_1=a/(1-c)$ and $b_1=b/(1-c)$. In these two particular cases, $C_{0}$ and $C_{1}$ are two integration constants that may be provided in terms of $M$ and $R$. Consequently, the physical parameters for this solution are
\begin{eqnarray}
\frac{b^2C_0 r^2-2abC_0r^2+2(b-a)(1-c)(C_0r^2+C_1)}{3(C_0r^2+C_1)^{\frac{b}{1-c}+2}(1-c)^2} &=& 8\pi p(r)\big[1-\alpha\{\rho(r)-3p(r)\}\big]\\
\frac{2bC_0(c-1)(3C_1+C_0r^2)-b^2C_0^2r^2}{B(C_0r^2+C_1)^{\frac{b}{1-c}+2}(1-c)^2} &=& 8\pi \rho(r)\big[1-\alpha\{\rho(r)-3p(r)\}\big]\\
\frac{b^2C_0r^2-2abC_0r^2+2(b-a)(1-c)(C_0r^2+C_1)}{2bC_0(c-1)(3C_1+C_0r^2)-b^2C_0^2r^2} &=& \frac{p(r)}{\rho(r)}.
\end{eqnarray}

\begin{figure}
\includegraphics[width=0.45\textwidth]{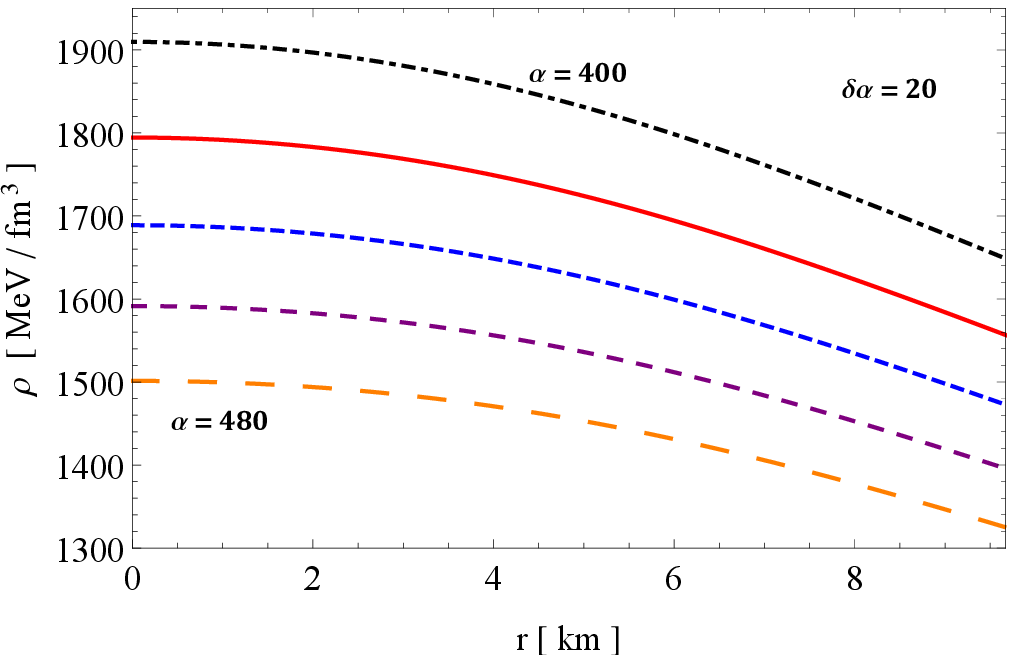}
\hspace{\fill}
\includegraphics[width=0.45\textwidth]{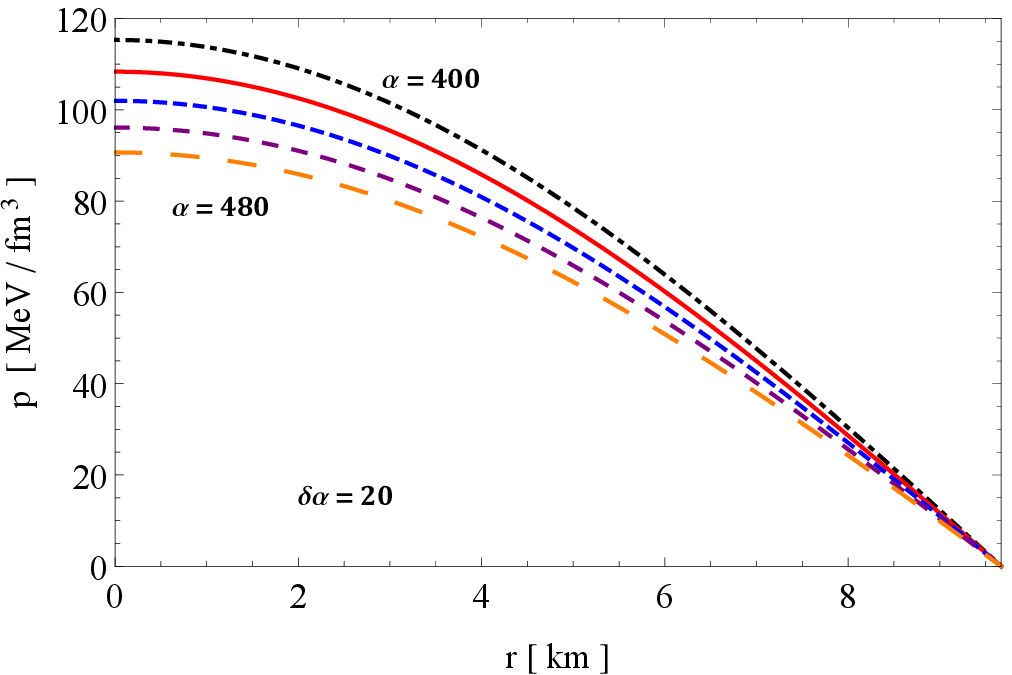}
\caption{Density and Pressure for case B.}\label{fig3}
\end{figure}

\begin{figure}
\includegraphics[width=0.45\textwidth]{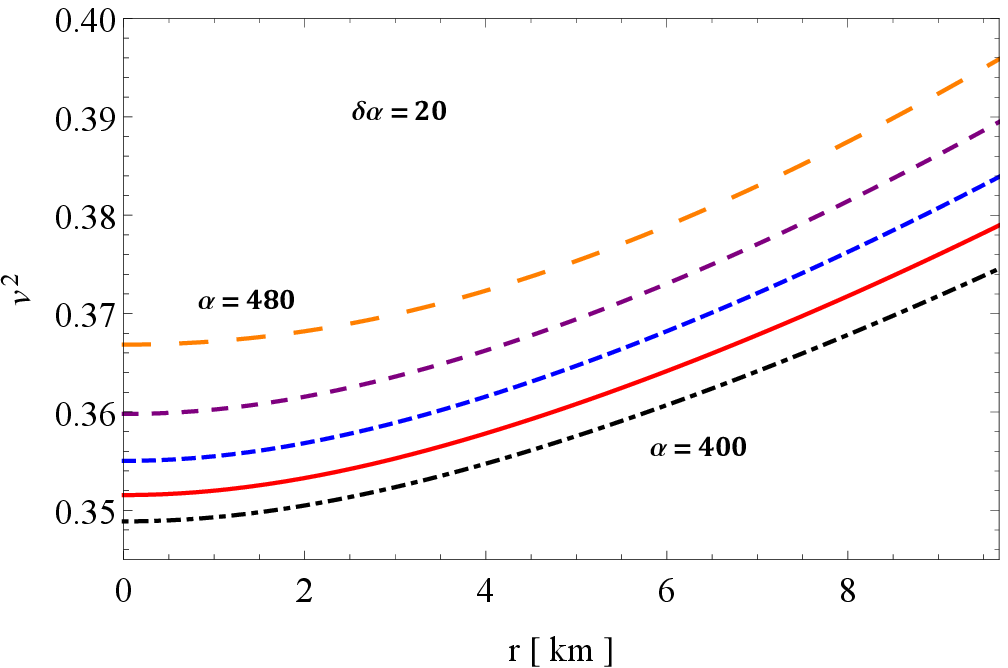}
\hspace{\fill}
\includegraphics[width=0.45\textwidth]{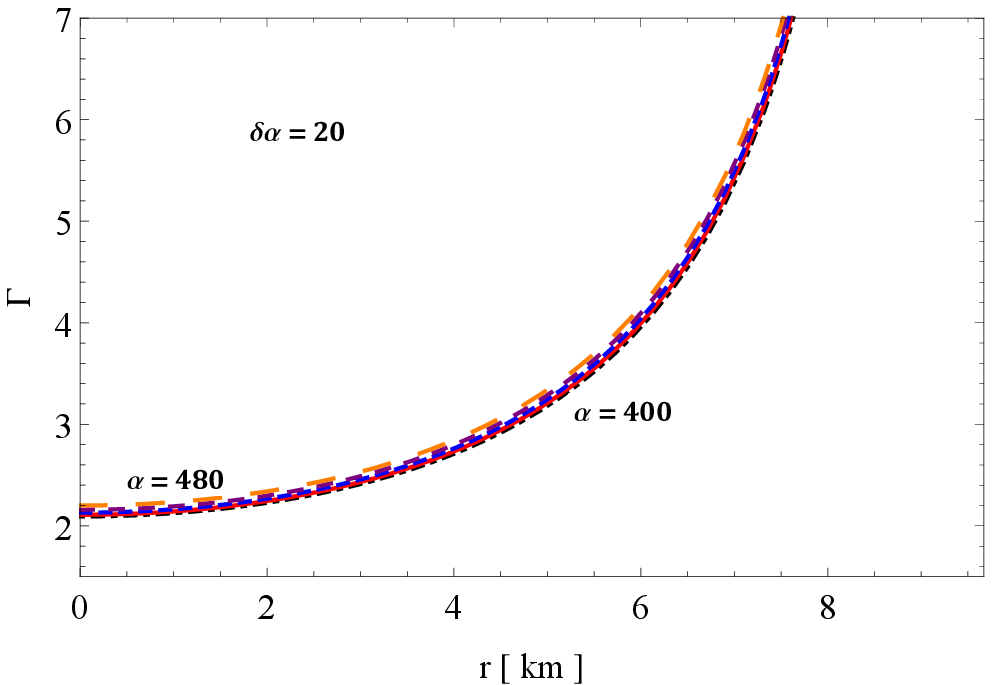}
\caption{Speed of sound and adiabatic index for case B.}\label{fig4}
\end{figure}

The variations of energy density and pressure can be seen in Figs. \ref{fig3}. Finally, the speed of sound and the adiabatic index can be determine as
\begin{eqnarray}
v^2 = {dp \over d\rho}~~;~~\Gamma={\rho+p \over p}\, {dp \over d\rho}.
\end{eqnarray}
These variations with respect to radial coordinates are shown in Fig. \ref{fig4}.

\section{Boundary Conditions}\label{sec.IV}
A spherically symmetric static fluid distribution described by the metric (\ref{isotropic}) should match with the exterior field described by the Schwarzschild solution given by

\begin{equation}
ds^{2}=\Big(1-\frac{2M}{r}\Big)dt^{2}-\Big(1-\frac{2M}{r}\Big)^{-1}dr^{2}-r^{2}(d\theta^{2}+\sin^{2}\theta \,d\phi^{2}\Big)
\end{equation}
Introducing the radial coordinate transformation
\begin{equation}
r=\tilde{r}\Big(1+\frac{M}{2\tilde{r}}\Big)^{2},
\end{equation}
the metric takes the following form
\begin{equation}
ds^{2}=\frac{\Big(1-{M^{2}/4\tilde{r}^{2}}\Big)^{2}}{\Big(1+{M/2\tilde{r}}\Big)^{4}}dt^{2}-\Big(1+\frac{M}{2\tilde{r}}\Big)^{4}(d\tilde{r}^{2}+\tilde{r}^{2}d\theta^{2}+\tilde{r}^{2}\sin^{2}\theta \,d\phi^{2})
\end{equation}
Thus, by matching the interior solution to the external vacuum solution on the boundary $\tilde{r}=R$ we get
\begin{eqnarray}
e^{\nu(R)}=\frac{\Big(1-{M^{2}/4R^{2}}\Big)^{2}}{\Big(1+{M/2R}\Big)^{4}}~~~ \mbox{and}~~~
e^{\mu(R)}=\Big(1+\frac{M}{2R}\Big)^{4}, \label{e39}
\end{eqnarray}
the quantity $R$ above represents the boundary of the star, where the pressure vanishes, and $M$ is the total gravitational mass of the star. This can be used to get an expression for $R$ in the next section

\subsection{For $c=1$:}
Using the matching condition \eqref{e39}, we can re-write as
\begin{eqnarray}
A_1 e^{-a_1 R^2} &=&\frac{\Big(1-{M^{2}/4R^{2}}\Big)^{2}}{\Big(1+{M/2R}\Big)^{4}}~~~;~~
B_1 e^{b_1 R^2} = \Big(1+{M \over 2R}\Big)^{4}.
\end{eqnarray}
This leads to 
\begin{eqnarray}
A_1 &=& e^{a_1 R^2} \frac{\Big(1-{M^{2}/4R^{2}}\Big)^{2}}{\Big(1+{M/2R}\Big)^{4}}~~;~~B_1=e^{-b_1 R^2}\Big(1+{M \over 2R}\Big)^{4},
\end{eqnarray}
and the vanishing pressure at the boundary $r=R$ gives
\begin{eqnarray}
a_1=\frac{b_1 \left(b_1 R^2+2\right)}{2 b_1 R^2+2}.
\end{eqnarray}
Further, we have chosen $M,R$ and $b_1$ as free parameters. 

\subsection{For $c \neq 1$:}
Similarly,  the matching condition in \eqref{e39} can be re-write as
\begin{eqnarray}
A \left(C_0 R^2+C_1\right)^{a_1} &=&\frac{\Big(1-{M^{2}/4R^{2}}\Big)^{2}}{\Big(1+{M/2R}\Big)^{4}}~~~;~~
B \left(C_0 R^2+C_1\right)^{b_1} = \Big(1+{M \over 2R}\Big)^{4}.
\end{eqnarray}
This leads to 
\begin{eqnarray}
A &=& \frac{(M-2 R)^2 \left(C_0 R^2+C_1\right)^{-a_1}}{(M+2 R)^2}~~;~~B=\frac{(M+2 R)^4 \left(C_0 R^2+C_1\right)^{-b_1}}{16 R^4},
\end{eqnarray}
and the vanishing pressure at the boundary $r=R$ gives
\begin{eqnarray}
a_1=-\frac{b_1 \left[(b_1+2) C_0 R^2+2 C_1\right]}{2 \left[(b_1+1) C_0 R^2+C_1\right]}.
\end{eqnarray}

Further, we have taken $M,~R,~b_1,~C_0$ and $C_1$ as free parameters. 
\section{Physical acceptability of the solution}\label{sec.V}

\begin{figure}
\includegraphics[width=0.6\textwidth]{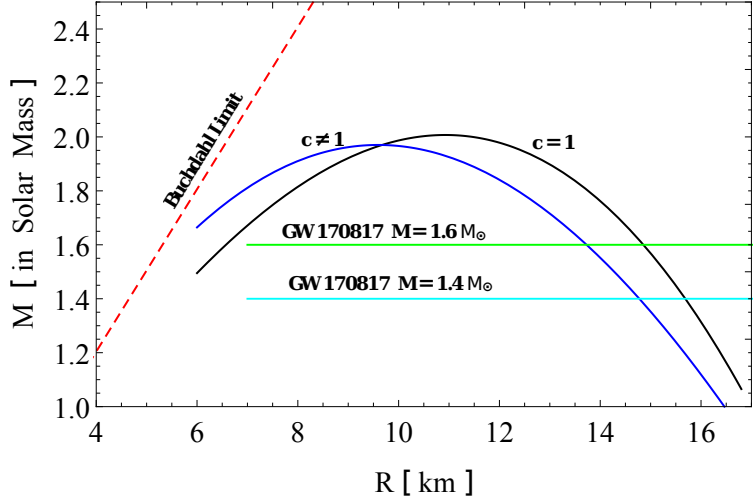}
\caption{$M-R$ curves for the solutions case A and B.}
\label{fig5}
\end{figure}

\begin{itemize}
  
\item[(a)]  The density $\rho$ is finite positive at $r=0$ and non-increasing towards the stellar surface, i.e. $d\rho/dr\leq0$.

\item[(b)]  The pressure is finite positive at $r=0$ and vanishes at the stellar surface i.e, $p(R)=0$.

\item[(c)] At the very center, it satisfies the Harrison-Zeldovich-Novikov criterion \cite{Harr65, zel71} i.e. 
\begin{eqnarray}
{p(0) \over \rho(0)} &=& {3b \over a-b} \leq 1 \Rightarrow 4b<a, ~~~~~~~~~~~~~~~~~~~~~~\mbox{(Case A)} \\
{p(0) \over \rho(0)} &=& {(b-a)(1-c) \over 3bC_0(c-1)} \leq 1 \Rightarrow 1+3C_0 \leq {a \over b}. ~~~\mbox{(Case B)}
\end{eqnarray}

\item[(d)]  The isotropic star satisfy the following energy conditions (see Figs. \ref{fig1} and \ref{fig3}):

1. Null energy condition (NEC): $\rho>0$

2. Weak energy condition (WEC): $p+\rho>0$

3. Strong energy condition (SEC): $\rho+3p>0$

4. Dominant energy condition (DEC): $\rho-p>0$

\item[(e)]  The solution holds the causality condition i.e., $1\geq dp/d\rho \geq 0$ (see Figs. \ref{fig2} and \ref{fig4}). 

\item[(f)]  The solution also holds the Bondi's criterion \cite{bondi64} i.e. $\Gamma \geq 4/3$ (see Figs. \ref{fig2} and \ref{fig4}). Hence, the solution is non-collapsing.

\item[(g)] The solution satisfies Buchdahl's bound \cite{Buch59} (see Fig. \ref{fig5}).

\end{itemize}

\section{Mass-radius relationship}\label{sec.VI}
For these two solutions we have generated the $M-R$ curves which can be seen in Fig. \ref{fig5}. Here one can see that the solution with $c=1$ can hold more massive object with larger radius ($M_{max}=2.00M_\odot,~R=10.95km$) than the case B solution ($c\neq 1$, $M_{max}=1.97M_\odot,~R=9.41km$). Further, we have incorporated the GW 170817 observational constraints that neutron stars of mass $1.6M_\odot$ and $1.4M_\odot$ should radius more that $10.68^{+0.15}_{-0.04}km$ \cite{bau17} and $11.0^{+0.9}_{-0.6}km$ \cite{cap20} respectively. From our solution, the two neutron star of mass $M_{max}=1.6M_\odot$ has radius of $13.73km$ (Case B) and $14.85km$ (case A), and for mass $1.4M_\odot$ has radius $14.77km$ (Case B) and $15.68km$ (case A). Hence, the solution satisfy the observed constraints from the neutron star merger event GW 170817. Hence, we can conclude that these first interior solutions in $\kappa(\mathcal{R},\mathcal{T})-$gravity are physically inspired. From Fig. \ref{fig5}, it can also be seen that the ratio of mass to radius is less than the Buchdahl limit, signifying that the solutions are non-collapsing. The $M-R$ curves generated from these solutions are almost similar to that of polytropic/SLy4 parameterization \cite{Cha97, Cha98,Art17}.
%%%%%%%%%%%%%%%%%%%%%%%
%\footnotesize
\begin{table*}
\centering
\caption{Central density, surface density, central pressure and predicted radii for neutron stars of masses $1.6M_\odot$ and $1.4M_\odot$.}\label{tab1}
\scalebox{0.8}{\begin{tabular}{|*{12}{c|} }
\hline
\multicolumn{6}{|c|}{{Case A}} & \multicolumn{6}{c|}{{Case B}} \\
\hline
\multirow{2}{*}{$\alpha$} & $\rho_c$ & $\rho_s$ & $p_c$ & \multirow{2}{*}{$R_{1.6}$} & \multirow{2}{*}{$R_{1.4}$} & \multirow{2}{*}{$\alpha$} & $\rho_c$ & $\rho_s$ & $p_c$ & \multirow{2}{*}{$R_{1.6}$} & \multirow{2}{*}{$R_{1.4}$} \\
\cline{2-4} \cline{8-10} 
$km^2$ & \multicolumn{3}{c|}{{$MeV/fm^3$}} & $km$ & $km$ & $km^2$ & \multicolumn{3}{c|}{{$MeV/fm^3$}} & $km$ & $km$\\
\hline
500 & 1515.77 & 1273.65 & 58.07 & & & 400 & 1908.95 & 1699.85 & 115.21 & & \\
\cline{1-4} \cline{7-10} 
520 & 1447.93 & 1213.99 & 55.54 & & & 420 & 1794.53 & 1556.64 & 108.34 & & \\
\cline{1-4} \cline{7-10} 
540 & 1387.11 & 1157.15 & 53.14 & 14.85 & 15.68 & 440 & 1689.13 & 1473.83 & 101.52 & 13.73 & 14.77 \\
\cline{1-4} \cline{7-10} 
560 & 1328.62 & 1106.38 & 51.00 & & & 460 & 1591.26 & 1395.53 & 95.77 & & \\
\cline{1-4} \cline{7-10} 
580 & 1274.82 & 1060.76 & 48.87 & & & 480 & 1500.93 & 1324.77 & 90.29 & & \\
\hline 
\end{tabular}}
\end{table*}

\section{Conclusions}\label{sec.VII}
We have successfully presented two exact interior solutions for the first time in $\kappa(\mathcal{R},\mathcal{T})$ gravity. The field equation has been derived by assuming isotropic coordinates in Schwarzschild's form. To solve the field equations, we have chosen the running gravitational constant as $\kappa(\mathcal{R},\mathcal{T})=8\pi-\lambda \mathcal{T}$ along with the isotropic pressure. We have then obtained two solutions for $c=1$ and $c\neq 1$. The case A solution is in the Krori-Barua form and the case B is more like the Tolman-Finch-Skea form. To obtain the constant  parameters, we have matched the interior solution with exterior Schwarzschild spacetime in isotropic form along with the condition $p(R)=0$. We have used the mass $M$ and radius $R$ as input observational parameters along with a few other constants. Further, we have plotted the variations of density, pressure, speed of sound, and adiabatic index to see the thermodynamic nature of these solutions. Hence, we found that these solutions are non-singular at $r=0$, non-increasing tends of density and pressure, obeyed causality condition, and satisfied Bondi's criterion. This further implies that the solutions also satisfy the energy conditions which can represent the non-collapsing stellar system. The central values of density \& pressure, surface density and predicted radii of $1.6M_\odot$ and $1.4M_\odot$ neutron stars are given in Table \ref{tab1}. Further, we have also shown that these solutions satisfy the constraints on mass and radius of neutron stars provided by the observation of the gravitational waves from the neutron star merger GW 170817. Hence, we conclude that these solutions are physically inspired.

\subsection*{Acknowledgments}
Farook Rahaman would like to thank the authorities of the Inter-University Centre for Astronomy and Astrophysics, Pune, India for providing the research facilities.

%%%%%%%%%%%%%%%%%%%%%%%%%%%%%%%%%%%%%%%%%%%%

\end{document}